%% file: main.tex
\documentclass[11pt]{article}
\usepackage[preprint]{acl}
\usepackage{times}
\usepackage{latexsym}
\usepackage[T1]{fontenc}
\usepackage[utf8]{inputenc}
\usepackage{microtype}
\usepackage{inconsolata}
\usepackage{graphicx}
\usepackage{amsmath,amssymb,booktabs,tabularx}
\usepackage{xcolor}
\newif\ifshowablationtodos
\showablationtodostrue

\newcommand{\method}{S3-S2ST}

\newcommand{\methodoffline}{\method}

\title{S3-S2ST: Semantically Supervised Streaming Speech-to-Speech Translation}
\author{
 \textbf{Yuxiang Zhao\textsuperscript{1,2,*}},
 \textbf{Zhiyun Fan\textsuperscript{3,*}},
 \textbf{Yu Pu\textsuperscript{3}},
  \textbf{Yanjie An\textsuperscript{1}},
 \textbf{Yushen Chen\textsuperscript{1,2}},
 \textbf{Wenrui Liang\textsuperscript{3}},
 \textbf{Baiji Liu\textsuperscript{3}},\\
 \textbf{Yunchong Xiao\textsuperscript{3}},
 \textbf{Zhaoci Liu\textsuperscript{3}},
 \textbf{Keqi Deng\textsuperscript{4}},
 \textbf{Kai Yu\textsuperscript{1}},
 \textbf{Zhifu Gao\textsuperscript{3}},
 \textbf{Xiangang Li\textsuperscript{3}},
 \textbf{Xie Chen\textsuperscript{1,2,$\dagger$}},
\\
\\
 \textsuperscript{1}X-LANCE Lab, School of Computer Science, Shanghai Jiao Tong University,
 \\
 \textsuperscript{2}Shanghai Innovation Institute,
 \textsuperscript{3}Alibaba Token Foundry,
 \textsuperscript{4}Independent Researcher,
\\
 \small{
   \textsuperscript{*}Equal contribution.\quad \textsuperscript{$\dagger$}Corresponding author.}
 \\
 \small{\textbf{Correspondence:} \href{mailto:yuxiangzhao@sjtu.edu.cn}{yuxiangzhao@sjtu.edu.cn}
 }
}
\begin{document}
\maketitle
\begin{abstract}
\input{text/00abstract}

\end{abstract}
\input{text/01introduction}
\input{text/02related_work}
\input{text/03method}
\input{text/04experimental_setup}
\input{text/05results}

\input{text/06ablations}
\input{text/07conclusion}
\input{text/08limitations}

\bibliography{references}
\clearpage
\appendix
\input{text/10implementation}
\end{document}

%% file: text/00abstract.tex
Streaming speech-to-speech translation requires timely decisions about when partial input is sufficient for accurate translation. We present S3-S2ST, a model that learns to translate and speak from semantically aligned streaming supervision. By converting utterance-level speech pairs into incremental training examples, we teach a learned policy when to commit translations. Its adjustable threshold controls the quality--latency tradeoff on a fixed 640\,ms input grid. S3-S2ST delivers competitive English--Chinese translation on FLEURS and leads the evaluated streaming systems in translation quality on the internal 9-pair test set. On RealSI, adapting the threshold to buffered speech further improves quality over a low fixed threshold with little change in first-audio latency. We have released the training and inference code and 4B model weights at \url{https://github.com/zhaoyx239/S3-S2ST}.

%% file: text/01introduction.tex
\section{Introduction}
\input{figure/01_application}
Streaming speech-to-speech translation (S2ST) enables conversation across languages by producing translated speech before the speaker finishes. Its central challenge is deciding how much context to wait for: early responses reduce delay but risk mistranslation, while longer waits interrupt the interaction. Unlike displayed text, speech already heard cannot be silently revised. A useful system must therefore coordinate translation quality, commitment timing, and speech generation (Figure~\ref{fig:application}).

Existing approaches address this challenge through different forms of coordination. Cascaded systems such as X-Translator reuse strong recognition, translation, and synthesis components, but must manage partial hypotheses and commitments across their interfaces \citep{zhao2026xtranslator}. Direct streaming models instead learn this coordination through monotonic attention or auxiliary alignment tasks, as in SeamlessStreaming and StreamSpeech \citep{seamless2023,zhang-etal-2024-streamspeech}. More recently, speech language models have brought stronger pretrained translation and generation capabilities: SimulS2S-LLM adapts offline models for simultaneous inference, while Hibiki and SimulS2ST-Omni learn from aligned speech streams \citep{deng-etal-2025-simuls2s,labiausse2025hibiki,he2026simuls2st}.

These advances motivate two closely related questions: how to supervise translation from incomplete speech, and how to adjust waiting behavior at inference. Utterance-level pairs specify the final translation but not which partial inputs support each target span. Streaming supervision must make that relationship explicit. Latency control also matters: SimulS2ST-Omni obtains different operating points by varying input chunk duration \citep{he2026simuls2st}. Larger chunks provide more context but reduce opportunities to respond to newly arriving evidence. This motivates controlling the decision to translate independently of input granularity.

We present \method{} (Semantically Supervised Streaming), which learns incremental translation and commitment from semantically aligned speech pairs. Our approach connects supervision for when to speak with continuous speech generation in a single streaming model. A tunable policy changes how much evidence is required before translating while preserving frequent decisions; playback adaptation further uses buffered speech to wait for useful context without delaying the initial commitment. This provides different quality--latency operating points from one checkpoint, without retraining.

Our contributions are threefold:
\begin{enumerate}
\item \textbf{Semantically supervised streaming.} We turn utterance-level speech pairs into supervision for what to translate and when to speak, combining chunkwise encoding with continuous Thinker--Talker conditioning.
\item \textbf{Flexible latency control.} Fixed and playback-adaptive thresholds control commitment on an unchanged 640\,ms input grid, enabling quality--latency adjustment without retraining.
\item \textbf{A strong streaming base model.} \method{} achieves competitive quality on FLEURS and leads the evaluated streaming baselines in translation COMET on internal 9-pair, with favorable quality--latency results on RealSI.
\end{enumerate}

%% file: figure/01_application.tex
\begin{figure*}[t]
\centering
\includegraphics[width=\textwidth]{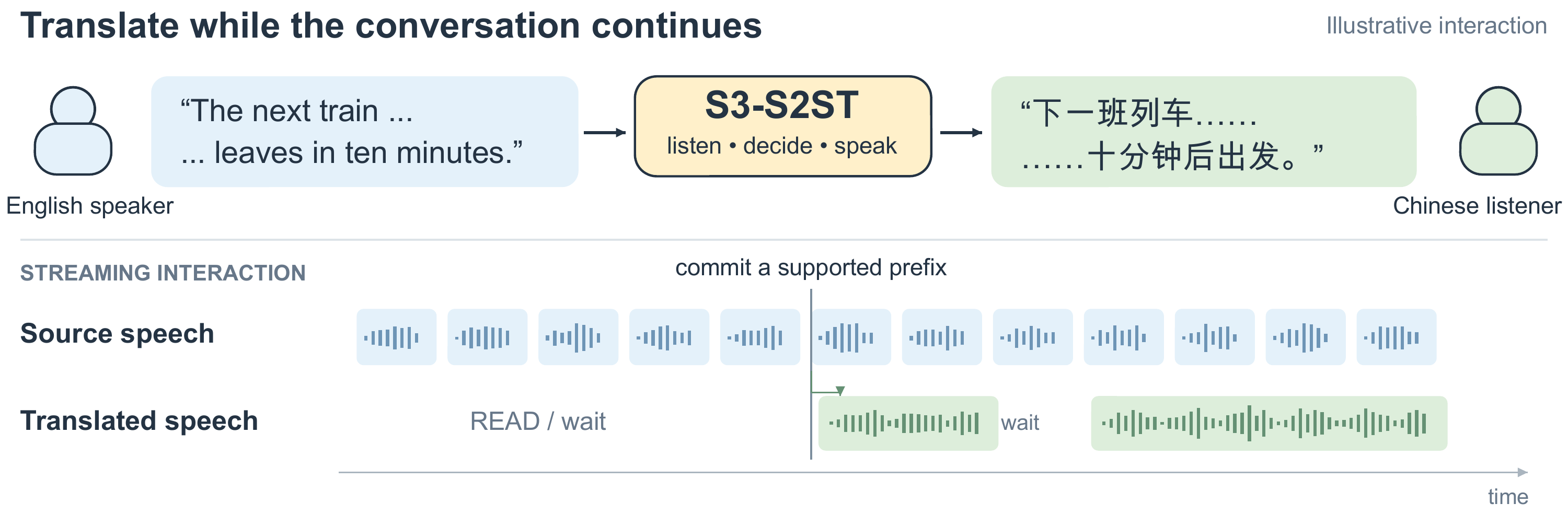}
\caption{Application overview of \method{}. The system listens to partial source speech, decides when to commit, and produces translated speech while the source continues. The timeline is illustrative, not a measured latency trace.}
\label{fig:application}
\end{figure*}

%% file: text/02related_work.tex
\section{Related Work}
\subsection{Direct and expressive S2ST.}
S2UT predicts discrete speech units directly \citep{lee-etal-2022-direct}, while UnitY uses translated text representations to guide speech generation \citep{inaguma-etal-2023-unity}. The Seamless family extends this direction to multilingual, expressive, and streaming translation \citep{seamless2023}. Thinker--Talker models such as Qwen2.5-Omni and Qwen3-Omni connect language modeling with speech generation \citep{xu2025omni,xu2025qwen3omni}, and UniSS emphasizes expressiveness and voice preservation \citep{cheng2025uniss}. DuraS2ST targets duration alignment through supervised initialization and reinforcement learning \citep{deng2026duras2st}. X-Dubbing couples time-aware translation with video-aware speech synthesis for synchronized dubbing \citep{choi2026xdubbing}. These advances improve the content and delivery of translated speech, but fluent generation alone does not determine when incomplete source speech provides enough context to translate.

\subsection{Streaming policies and supervision.}
Wait-$k$ offers simple, predictable latency control through a fixed schedule \citep{ma-etal-2019-stacl}, but cannot adapt its waiting decisions to individual inputs. Adaptive segmentation and DiSeg learn translation boundaries \citep{zhang-etal-2022-learning,zhang-feng-2023-end}; EMMA and StreamSpeech use monotonic attention and auxiliary alignment tasks, respectively, to coordinate source consumption and output \citep{ma2023emma,zhang-etal-2024-streamspeech}. SimulS2S-LLM enables simultaneous inference from offline training through boundary-aware prompts and a test-time policy \citep{deng-etal-2025-simuls2s}. Hibiki and SimulS2ST-Omni instead learn coordinated speech streams from aligned supervision \citep{labiausse2025hibiki,he2026simuls2st}, which specifies timing information absent from utterance-level corpora such as CVSS \citep{jia-etal-2022-cvss}. Hibiki-Zero removes word-level alignment supervision and uses reinforcement learning to optimize latency while preserving translation quality \citep{labiausse2026hibikizero}. X-Translator reuses pretrained recognition, translation, and synthesis modules, coordinating their incremental outputs through explicit interfaces \citep{zhao2026xtranslator}.

These approaches differ in how they expose latency control. SimulS2ST-Omni varies input chunk duration to obtain latency tiers: larger chunks supply more context but reduce decision frequency, without forcing fixed output segments \citep{he2026simuls2st}. DIAL-S2ST learns latency control through lookahead self-distillation and adjusts the scale of a single LoRA adapter at inference \citep{choi2026dials2st}. This controls emission timing through effective model weights. Our explicit READ/WRITE threshold can instead vary within an utterance according to the playback buffer, without changing model weights or input granularity.

\subsection{Speech conditioning and evaluation.}
Learned layer mixtures combine representations from different model depths \citep{peters-etal-2018-deep}; SimulS2S-LLM applies multi-layer conditioning to speech-token prediction \citep{deng-etal-2025-simuls2s}. Evaluating the resulting speech requires more than text translation scores. OpenSTBench covers semantic, acoustic, and temporal quality \citep{an2026openstbench}, while SimulEval standardizes simultaneous evaluation \citep{ma-etal-2020-simuleval}. Studies of continuous streaming and computation-aware latency show why evaluation must account for online processing as well as source consumption \citep{papi-etal-2025-real,xu-etal-2025-ca}. Together, these protocols support assessing translation quality alongside the delay experienced by listeners.

%% file: text/03method.tex
\section{Method}
\input{figure/02_architecture}
\subsection{Model Interfaces}
Let $x_{1:T}$ be source speech, $y_{1:N}$ target text, and $z_{1:M}$ target codec tokens. The Thinker incrementally generates $y$; the Talker generates $z$ from Thinker states, and a frozen flow-matching decoder and vocoder render speech (Figure~\ref{fig:architecture}). The streaming context interleaves source-audio and translated-text spans. Semantic boundaries provide READ/WRITE supervision during training; an auxiliary classifier predicts these decisions at inference. The offline model receives the complete source before translation.

\subsection{640 ms Chunkwise Encoder}
The frontend advances in 640\,ms chunks of 16 encoder states; a block-causal adaptor reduces each chunk to eight positions. It allows within-chunk right context and 160\,ms of lookahead at the first encoder layer. Later layers cannot attend to the next block, preventing successive layers from extending the future-context budget.

We retain pretrained weights and subsampling convolutions. The 31-tap FSMN convolutions preserve centered tap alignment and mask unavailable right context, rather than shifting all taps into the past. Inference uses only received audio, waiting for the declared lookahead before processing a block. The 640\,ms block defines observation granularity, while actual audio-response latency also includes generation and rendering. Appendix~\ref{app:model} specifies the masks and frontend implementation.

\subsection{Predicting When to Translate}
At each complete acoustic interval, an MLP maps the final normalized Thinker state to a WRITE probability $p_t$. Cross-entropy with equal class weights trains the gate on READ/WRITE targets. Inference selects WRITE when $p_t\geq\tau$; end-of-input forces a residual flush. Raising $\tau$ requires stronger evidence without updating weights, although the changed generation history means quality need not improve monotonically.

The playback-adaptive policy uses pending playback to regulate commitment. Let $b_t\geq0$ be the playable audio remaining in the output buffer, in seconds. We set
\begin{equation}
\tau_t=0.1+0.89\sqrt{\min\!\left(\frac{b_t}{1\,\mathrm{s}},1\right)}.
\label{eq:dynamic-threshold}
\end{equation}The adaptive threshold ranges from $0.1$ to $0.99$. 
An empty buffer restores $\tau_t=0.1$, allowing an early response. While speech remains available for playback, the higher threshold lets the model collect more source context before committing. The rule changes inference decisions without additional training; end-of-input still forces a flush.

\subsection{Eight-Layer Hidden-State Fusion}
For the 4B model, the Talker receives states from $\mathcal{L}=\{4,8,12,16,20,24,28,32\}$. A learnable vector $w_\ell\in\mathbb{R}^{2560}$ weights each layer channelwise before the existing projection $P$:
\begin{equation}
\widetilde h_t=\sum_{\ell\in\mathcal{L}}w_\ell\odot h_t^{(\ell)},\qquad q_t=P\widetilde h_t.
\label{eq:fusion}
\end{equation}
The weights are learned jointly with the conditioning interface and shared across positions. Text and gate heads continue to use the final Thinker state. Appendix~\ref{app:model} specifies the coefficient parameterization and state conventions.

\subsection{Training Data and Supervision}
\label{sec:data}
\subsubsection{Paired speech for offline training.}The source ASR corpora comprise public datasets and data obtained through licensed procurement. 
We construct bilingual pairs from ASR data: Qwen3-Max translates the source transcript, and CosyVoice3.5 synthesizes target speech. English-to-Chinese and Chinese-to-English examples each use real recordings in the source language, without reversing synthetic targets into additional sources. The resulting source audio, target text, and target codec sequences support offline speech adaptation, where the complete source is available before target generation. This stage learns speech realization and the Thinker--Talker interface, and supplies the speech components for streaming training.

\subsubsection{Additional supervision for streaming.}
Qwen3.7-Max divides paired transcripts into aligned semantic units. Direction-specific prompts enforce prefix-supported translation, and validation requires exact transcript reconstruction (Appendix~\ref{app:semantic-labeling}). Forced alignment locates each unit in both waveforms; filtering removes invalid spans, nonmonotonic timestamps, and inconsistent durations. A source endpoint $a_j$ is quantized to $g_j=C\lceil a_j/C\rceil$, where $C=0.64$\,s. Both text and speech adaptation use these grid-aligned boundaries. Target speech is mapped to a 12.5\,Hz codec timeline.

Each unit becomes a trajectory: consume source audio, supervise READ at internal grid points, supervise WRITE at the unit boundary, then predict the target text and codec segment. A partial final span is flushed at end-of-input. Units can span different numbers of blocks, linking commitment to semantic evidence while retaining a fixed observation grid.

The annotator uses complete transcript pairs to construct training targets; at inference, the model predicts commitment from received source prefixes. Section~\ref{sec:setup} specifies the separate offline and streaming data mixtures and training schedules.

\subsection{Speech Generation and Training}
Continuous conditioning retains audio and text states in temporal order without interior padding. After separate text and speech adaptation, joint streaming training updates the encoder, adaptor, Thinker, gate, fusion, projection, and Talker using
\begin{equation}
\mathcal L=\mathcal L_{\mathrm{text}}+\mathcal L_{\mathrm{codec}}+0.1\mathcal L_{\mathrm{gate}}.
\end{equation}
The waveform renderer remains frozen: end-to-end training refers to source-to-codec optimization. Section~\ref{sec:setup} describes the training stages; Appendix~\ref{app:model} provides implementation details.

%% file: figure/02_architecture.tex
\begin{figure*}[t]
\centering
\includegraphics[width=\textwidth]{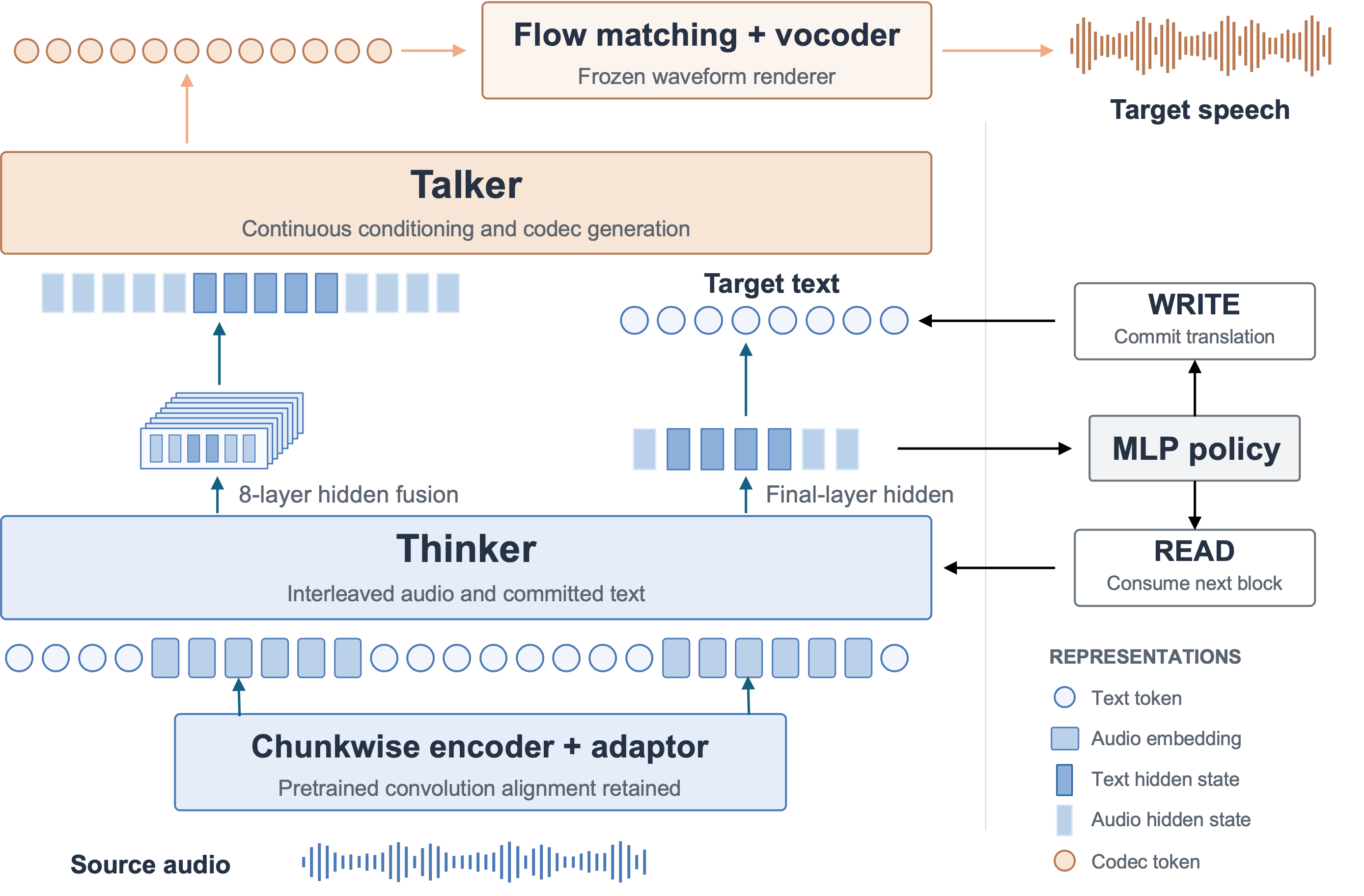}
\caption{Architecture of \method{}. Chunkwise acoustic features and committed text form the Thinker context. Its final state supports text generation and an MLP READ/WRITE policy; a separate weighted fusion of eight layers conditions the Talker through a projection. Codec tokens are rendered by frozen flow matching and vocoder modules. READ advances the source; WRITE enables generation of a translation span. Token counts are schematic. Trainability of the other modules depends on the training stage.}
\label{fig:architecture}
\end{figure*}

%% file: text/04experimental_setup.tex
\section{Experimental Setup}
\label{sec:setup}
\subsection{Training Data}
\subsubsection{Streaming speech data.}
The final streaming mixture contains approximately 13k hours of real source audio, with roughly 6.5k hours per translation direction. We filter both directions by source and target ASR consistency and balance them by source duration. The pairs follow the construction procedure in Section~\ref{sec:data}. The mixture also includes approximately 10 hours of retained FLEURS training speech, upsampled tenfold.

\subsubsection{Offline speech data.}
Offline speech adaptation uses approximately 300k hours of source speech, including upsampled FLEURS training data in its continuation segment. Both final speech-training mixtures exclude the added 9-pair training supplement; the preceding text-adaptation stage includes separate 9-pair training sources. Exact-match checks find no overlap with the evaluation manifests. Appendix~\ref{sec:implementation} gives the mixture composition, filtering thresholds, and audit scope.

% Submit the wide result float early for placement near the results.
\input{table/01_s2st_offline}

\subsection{Training Stages}
For the 4B model, training has three stages: streaming S2TT adaptation, offline S2ST adaptation, and streaming S2ST adaptation. S2TT first adapts the frontend, then learns incremental translation and commitments. Offline S2ST supplies speech-generation components. Streaming S2ST combines both branches and trains them through Talker and E2E substages. The 35B/3B variants follow the same staged training recipe as the 4B models.

\paragraph{Streaming S2TT adaptation.}
We first train the chunkwise encoder, adaptor, and Thinker LoRA for 10,000 updates on the original utterance-level S2TT corpus (approximately 3M hours), with an effective batch of 128 packed sequences. After merging LoRA weights, we train the Thinker and READ/WRITE gate for another 10,000 updates on approximately 17k hours of semantic trajectories, freezing the encoder and adaptor and using an effective batch of eight.

\paragraph{Offline S2ST adaptation.}
With the source frontend and Thinker frozen, we train the Talker, codec embeddings/output head, projection, and fusion under one objective for 15,000 updates: 5,000 initial updates and 10,000 continuation updates on the expanded offline pool. Continuation starts a fresh optimizer and scheduler while retaining the same trainable modules. Both segments use an effective batch of 64. The resulting checkpoint is the offline \method{} model and supplies the speech components for streaming initialization.

\paragraph{Streaming S2ST adaptation.}
We combine the S2TT frontend, Thinker, and gate with the offline speech components. The Talker substage adapts the speech-generation components for 5,000 updates with the translation components frozen; the E2E substage then jointly optimizes source-to-codec translation. The reported streaming checkpoint uses 4,000 E2E updates. Both substages use the same filtered streaming mixture and effective batch of 64. The waveform renderer remains frozen throughout. Optimizer and warmup settings are in Appendix~\ref{app:training}; packing and numerical details are in Appendix~\ref{app:model}.

\subsection{Inference}
For the 4B models, the streaming variant consumes 640\,ms source blocks with 160\,ms lookahead. The MLP selects READ or WRITE from the interleaved audio--text context. Committed text is replayed to recover speech-conditioning states; fused states continuously condition codec generation. The Thinker uses greedy decoding (temperature $0$, top-$k=1$; nucleus sampling inactive). Both 4B S2ST models use Talker temperature $0.8$, top-$k=5$, top-$p=0.8$, and seed $0$.

Streaming uses eight-token flow-matching chunks (FM8). The main tables report both fixed $\tau=0.5$ and the playback-adaptive threshold in Equation~\ref{eq:dynamic-threshold}; additional fixed thresholds characterize the quality--latency curve. Speaker conditioning is refreshed at each nonempty WRITE from all source speech consumed so far, without reading future audio. The offline model uses up to ten seconds of fully received source speech and FM15 rendering. We also evaluate 35B-total, 3B-active MoE variants in both offline and streaming settings on FLEURS and 9-pair (Tables~\ref{tab:s2st-offline} and~\ref{tab:s2st}). 

\input{table/02_s2st_streaming}
\subsection{Evaluation Data and Metrics}
We evaluate on the publicly available English--Chinese FLEURS test data, containing 945 Chinese-to-English and 646 English-to-Chinese examples \citep{conneau2022fleurs}. The internal 9-pair test set contains 996 examples per direction and provides an additional evaluation distribution; it is not publicly released. Each system is evaluated on the same manifest within each dataset, and reported scores are unweighted means of the two direction-level scores. All component ablations use FLEURS. We use the public RealSI benchmark\footnote{\url{https://github.com/byteresearchcla/RealSI}}~\citep{cheng2024clasi} for quality--latency evaluation, following the benchmark choice of \citet{he2026simuls2st}. Policy-development splits and evaluation details are specified in Appendices~\ref{sec:implementation} and~\ref{app:evaluation}.

Following OpenSTBench \citep{an2026openstbench}, we assess text quality, speech quality, and latency. We report corpus BLEU \citep{post-2018-call}, COMET using XCOMET-XL\footnote{\url{https://huggingface.co/Unbabel/XCOMET-XL}} \citep{rei-etal-2020-comet,guerreiro-etal-2024-xcomet}, and ASR-BLEU and ASR-COMET on transcribed output speech. WER compares ASR output with generated text to measure synthesis fidelity; it averages English WER and Chinese CER (percent). SIM is source--output ERes2Net cosine similarity, and UTMOS estimates naturalness \citep{saeki2022utmos}.

Computation-aware StartOffset measures elapsed time from the start of real-time source replay to the first nonempty audio packet, including source waiting and online computation. We follow the audio-emission protocol of SimulEval/OpenSTBench \citep{ma-etal-2020-simuleval,an2026openstbench}; quality and latency use the same generated waveforms. Appendix~\ref{app:evaluation} details timing boundaries and scoring.

\subsection{Baselines}
Tables~\ref{tab:s2st-offline} and~\ref{tab:s2st} separate offline and streaming systems: SeamlessM4T-v2-large and SeamlessStreaming \citep{seamless2023}, Qwen3-Omni \citep{xu2025qwen3omni}, UniSS \citep{cheng2025uniss}, SimulS2ST-Omni \citep{he2026simuls2st}, X-Translator \citep{zhao2026xtranslator}, and text-only IWSLT (VAD) \citep{agostinelli-etal-2025-findings}. We additionally evaluate Hy-SimulMT\footnote{\url{https://github.com/Tencent-Hunyuan/Hy-SimulMT}} (30B-A3B-Interleave) with its official Qwen3-TTS (0.6B) backend: both native chunk settings on FLEURS and 9-pair (Table~\ref{tab:s2st}) and RealSI (Figure~\ref{fig:s2st-latency}). Params. lists backbone sizes, except that $\dagger$ marks model-level counts; MoE sizes are reported as total/active backbone parameters. Text-only IWSLT contributes text metrics only. Appendix~\ref{app:evaluation} records baseline decoding, preprocessing, parameter-count scope, and scorer configurations.

%% file: table/01_s2st_offline.tex
\begin{table*}[!t]
\centering\small
\setlength{\tabcolsep}{2pt}
\begin{tabular*}{\textwidth}{@{\extracolsep{\fill}}llrrrrrrr@{}}
\toprule
System & Params. & BLEU$\uparrow$ & COMET$\uparrow$ & WER$\downarrow$ & ASR-BLEU$\uparrow$ & ASR-COMET$\uparrow$ & SIM$\uparrow$ & UTMOS$\uparrow$ \\
\midrule
\multicolumn{9}{l}{\textit{FLEURS}} \\
SeamlessM4T-v2-large & 2.3B$^{\dagger}$ & 26.20 & 0.800 & 7.63 & 24.80 & 0.728 & 0.026 & 3.16 \\
Qwen3-Omni & 30B/3B & \underline{34.64} & \underline{0.907} & 2.62 & \underline{31.50} & 0.844 & 0.014 & \textbf{4.16} \\
UniSS (Quality) & 1.5B & 29.91 & 0.825 & 2.57 & 26.48 & 0.787 & 0.475 & 3.89 \\
UniSS (Performance) & 1.5B & 28.78 & 0.807 & \textbf{1.98} & 25.34 & 0.772 & 0.475 & 3.88 \\
\methodoffline{} & 4B & 34.59 & 0.902 & \underline{2.13} & 30.05 & \underline{0.847} & \textbf{0.606} & 4.01 \\
\methodoffline{} & 35B/3B & \textbf{38.54} & \textbf{0.927} & 2.58 & \textbf{32.59} & \textbf{0.858} & \underline{0.539} & \underline{4.09} \\
\midrule
\multicolumn{9}{l}{\textit{9-pair}} \\
SeamlessM4T-v2-large & 2.3B$^{\dagger}$ & 38.08 & 0.914 & 5.05 & 37.40 & 0.868 & 0.062 & 3.20 \\
Qwen3-Omni & 30B/3B & 56.40 & 0.974 & 1.08 & 53.15 & 0.952 & 0.066 & \underline{4.14} \\
UniSS (Quality) & 1.5B & 48.22 & 0.943 & 1.03 & 47.11 & 0.928 & \underline{0.538} & 3.88 \\
UniSS (Performance) & 1.5B & 47.89 & 0.939 & \textbf{0.85} & 46.90 & 0.925 & 0.537 & 3.87 \\
\methodoffline{} & 4B & \underline{64.79} & \underline{0.979} & \underline{0.95} & \underline{58.95} & \textbf{0.956} & \textbf{0.581} & 4.11 \\
\methodoffline{} & 35B/3B & \textbf{65.47} & \textbf{0.981} & 1.13 & \textbf{59.16} & \underline{0.954} & 0.509 & \textbf{4.21} \\
\bottomrule
\end{tabular*}
\caption{Offline S2ST results on FLEURS and 9-pair. Bold and underline mark the best and second-best scores.}
\label{tab:s2st-offline}
\end{table*}

%% file: table/02_s2st_streaming.tex
\begin{table*}[!t]
\centering\small
\setlength{\tabcolsep}{2pt}
\begin{tabular*}{\textwidth}{@{\extracolsep{\fill}}llrrrrrrr@{}}
\toprule
System & Params. & BLEU$\uparrow$ & COMET$\uparrow$ & WER$\downarrow$ & ASR-BLEU$\uparrow$ & ASR-COMET$\uparrow$ & SIM$\uparrow$ & UTMOS$\uparrow$ \\
\midrule
\multicolumn{9}{l}{\textit{FLEURS}} \\
IWSLT (VAD) & 2.3B$^{\dagger}$ & 21.10 & 0.653 & -- & -- & -- & -- & -- \\
SeamlessStreaming & 2.5B$^{\dagger}$ & 23.44 & 0.739 & 12.70 & 22.17 & 0.647 & 0.034 & 3.36 \\
X-Translator & 8B & 28.60 & 0.752 & 4.90 & 25.28 & 0.701 & \textbf{0.673} & 3.12 \\
SimulS2ST-Omni (m2) & 3B & 26.68 & 0.748 & 12.01 & 21.45 & 0.625 & 0.595 & 3.20 \\
Hy-SimulMT (1.6\,s) & 30B/3B & 32.51 & 0.866 & \textbf{2.08} & 28.36 & 0.809 & 0.400 & 3.70 \\
Hy-SimulMT (0.8\,s) & 30B/3B & 30.24 & 0.811 & \underline{2.52} & 26.68 & 0.761 & 0.394 & 3.62 \\
\method{} (fixed) & 4B & 31.14 & 0.857 & 4.46 & 26.55 & 0.785 & \underline{0.610} & \underline{3.76} \\
\method{} (adaptive) & 4B & 30.95 & 0.837 & 3.86 & 26.87 & 0.773 & 0.609 & 3.66 \\
\method{} (fixed) & 35B/3B & \textbf{37.38} & \textbf{0.908} & 3.09 & \textbf{32.11} & \textbf{0.839} & 0.608 & \textbf{3.79} \\
\method{} (adaptive) & 35B/3B & \underline{36.11} & \underline{0.886} & 3.70 & \underline{31.36} & \underline{0.817} & 0.608 & 3.67 \\
\midrule
\multicolumn{9}{l}{\textit{9-pair}} \\
IWSLT (VAD) & 2.3B$^{\dagger}$ & 32.65 & 0.854 & -- & -- & -- & -- & -- \\
SeamlessStreaming & 2.5B$^{\dagger}$ & 32.90 & 0.871 & 11.77 & 32.44 & 0.792 & 0.060 & 3.40 \\
X-Translator & 8B & 44.73 & 0.907 & 2.73 & 42.54 & 0.885 & \underline{0.630} & 3.39 \\
SimulS2ST-Omni (m2) & 3B & 47.18 & 0.940 & 8.28 & 40.36 & 0.840 & \textbf{0.661} & 3.33 \\
Hy-SimulMT (1.6\,s) & 30B/3B & 52.22 & 0.954 & \underline{1.24} & 50.05 & 0.926 & 0.562 & 3.71 \\
Hy-SimulMT (0.8\,s) & 30B/3B & 48.34 & 0.931 & 2.12 & 46.73 & 0.903 & 0.560 & 3.66 \\
\method{} (fixed) & 4B & \underline{60.51} & 0.965 & 1.54 & \underline{55.13} & 0.933 & 0.585 & \underline{3.78} \\
\method{} (adaptive) & 4B & 57.95 & 0.956 & 1.62 & 53.04 & 0.922 & 0.588 & 3.67 \\
\method{} (fixed) & 35B/3B & \textbf{60.77} & \textbf{0.975} & \textbf{1.09} & \textbf{55.36} & \textbf{0.948} & 0.580 & \textbf{3.91} \\
\method{} (adaptive) & 35B/3B & 57.51 & \underline{0.967} & 1.43 & 52.88 & \underline{0.935} & 0.585 & 3.77 \\
\bottomrule
\end{tabular*}
\caption{Streaming translation results on FLEURS and 9-pair. ``--'' denotes inapplicable speech metrics.}
\label{tab:s2st}
\end{table*}

%% file: text/05results.tex
\section{Results}
\subsection{Translation and Speech Quality}
Tables~\ref{tab:s2st-offline} and~\ref{tab:s2st} summarize results on FLEURS and the internal 9-pair test set. S3-S2ST achieves strong translation quality in both offline and streaming settings. Scaling to 35B/3B improves most translation metrics, while the 4B offline model retains higher speaker similarity.

With fixed $\tau=0.5$, the 35B/3B streaming model leads the evaluated streaming systems in text and ASR-based translation metrics on both datasets. The 4B dense model remains competitive with Hy-SimulMT's 30B-total, 3B-active MoE backbone. For both sizes, fixed $\tau=0.5$ yields higher COMET than playback-adaptive decoding; the 4B quality--latency tradeoff is analyzed below. Speech fidelity, naturalness, and speaker similarity show complementary tradeoffs across systems.

\subsection{Quality and Latency}
\input{figure/03_s2st_tradeoff}
Figure~\ref{fig:s2st-latency} evaluates all systems on RealSI under a common timing protocol. Varying the fixed WRITE threshold traces the tradeoff between ASR-COMET and computation-aware StartOffset without changing the checkpoint or the 640\,ms input grid. Fixed $\tau=0.1$ improves both ASR-COMET and StartOffset over SimulS2ST-Omni m2--m5; m1 responds earlier with lower quality. The playback-adaptive point improves on fixed $\tau=0.1$ at essentially unchanged onset and approaches the quality of fixed $\tau=0.2$ with an earlier response. The adaptive point also achieves higher ASR-COMET and lower StartOffset than both Hy-SimulMT chunk settings. This supports using available playback time to gather context for later commitments while preserving an early first response. Table~\ref{tab:dynamic-threshold} provides the corresponding controlled comparison against fixed $\tau=0.1$ on FLEURS.

\subsection{Human Evaluation}
Table~\ref{tab:human} reports five-point ratings from 20 raters on 400 FLEURS utterances, with 200 per direction and 20 per rater. Scores are averaged over both directions with 95\% rater-cluster bootstrap intervals; protocol details appear in Appendix~\ref{app:human}. \method{} (4B, fixed $\tau=0.5$) receives the highest mean audio-quality rating and ranks second in translation quality and speaker similarity. Hy-SimulMT leads translation quality, while SimulS2ST-Omni leads speaker similarity.
\input{table/06_human_evaluation}

%% file: figure/03_s2st_tradeoff.tex
\begin{figure}[t]
\centering
\includegraphics[width=\columnwidth]{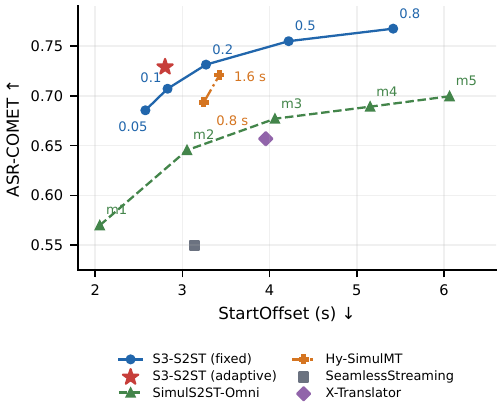}
\caption{Quality--latency tradeoff of S3-S2ST (4B) and baselines on RealSI, averaged equally across directions. Labels denote our fixed thresholds, SimulS2ST-Omni latency multipliers m1--m5, and Hy-SimulMT subsequent chunk durations. The star denotes our playback-adaptive policy.}
\label{fig:s2st-latency}
\end{figure}

%% file: table/06_human_evaluation.tex
\begin{table}[!htbp]
\centering\footnotesize
\setlength{\tabcolsep}{1.5pt}
\begin{tabularx}{\columnwidth}{@{}l*{3}{>{\centering\arraybackslash}X}@{}}
\toprule
System & \shortstack{Translation\\quality$\uparrow$} & \shortstack{Audio\\quality$\uparrow$} & \shortstack{Speaker\\similarity$\uparrow$} \\
\midrule
SeamlessStreaming & \mbox{$3.04\!\pm\!0.20$} & \mbox{$3.00\!\pm\!0.27$} & \mbox{$1.54\!\pm\!0.23$} \\
SimulS2ST-Omni (m2) & \mbox{$3.20\!\pm\!0.19$} & \mbox{$3.69\!\pm\!0.24$} & \mbox{$\mathbf{3.90}\!\pm\!0.21$} \\
X-Translator & \mbox{$3.33\!\pm\!0.16$} & \mbox{$3.79\!\pm\!0.21$} & \mbox{$3.65\!\pm\!0.21$} \\
Hy-SimulMT (1.6\,s) & \mbox{$\mathbf{3.99}\!\pm\!0.19$} & \mbox{$\underline{3.86}\!\pm\!0.19$} & \mbox{$3.52\!\pm\!0.24$} \\
\method{} (4B) & \mbox{$\underline{3.82}\!\pm\!0.21$} & \mbox{$\mathbf{4.06}\!\pm\!0.21$} & \mbox{$\underline{3.72}\!\pm\!0.19$} \\
\bottomrule
\end{tabularx}
\caption{Human scores on FLEURS, averaged over both directions: mean $\pm$ 95\% CI half-width.}
\label{tab:human}
\end{table}

%% file: text/06ablations.tex
\section{Ablations and Analysis}
\label{sec:ablations}All ablations use the 4B models on FLEURS; training and decoding details appear in Appendix~\ref{app:ablations}.

\subsection{Encoder Adaptation}
\input{table/03_encoder_comparison}
Table~\ref{tab:encoder-comparison} compares our chunkwise adaptation with the earlier block-causal recipe using its archived inference implementation, both at $\tau=0.5$. The chunkwise checkpoint follows streaming S2TT adaptation, before Talker and joint S2ST training, and thus differs from Table~\ref{tab:s2st}. It achieves stronger translation quality. This is a recipe-level comparison with different initialization and training settings (Appendix~\ref{app:ablations}).

\subsection{Commitment Policy}
\input{table/04_policy_ablation}
Table~\ref{tab:policy-ablation} compares the MLP with WRITEs every six or eight source blocks, keeping weights, FM8 rendering, and initial-prefix speaker conditioning fixed. The MLP improves quality over the eight-block schedule while responding earlier; six blocks give an earlier onset at a greater quality cost. These results favor content-dependent commitment over fixed intervals.

\subsection{Playback-Adaptive Threshold}
\input{table/07_dynamic_threshold}
Table~\ref{tab:dynamic-threshold} compares fixed and playback-adaptive thresholds using identical weights, FM8 rendering, and speaker conditioning. Adaptation improves ASR-COMET over $\tau=0.1$ with essentially unchanged first-audio latency and a modest increase in Average Token Delay (ATD)~\citep{kano2024atd}. It matches the quality and ATD of $\tau=0.2$ while producing the first audio 0.39\,s earlier.

\subsection{Speech-Conditioning Interface}
\input{table/05_fusion_ablation}
Table~\ref{tab:fusion-ablation} compares eight-layer fusion with final-layer conditioning using the same initialization sources, data, Talker and joint-training budgets, and decoding configuration. ASR-COMET measures translated speech content; WER measures fidelity to generated text. Eight-layer fusion modestly improves ASR-COMET and lowers WER, supporting intermediate representations for speech conditioning.
% Human evaluation retained for later restoration.
% \subsection{Human Evaluation}
% \input{table/06_human_evaluation}
% We plan a blinded listening evaluation on 60 FLEURS items, with 30 in each translation direction. Each item pairs the same source with outputs from S3-S2ST at fixed $\tau=0.5$, SeamlessStreaming, X-Translator, and SimulS2ST-Omni. System identities are hidden and presentation order is randomized. Twenty raters each assess nine items, giving three ratings per item and system.
% 
% Raters separately judge translation adequacy, audio naturalness, and source-speaker similarity on 1--5 scales. These dimensions distinguish whether the output preserves the message, sounds natural, and retains the source voice. Table~\ref{tab:human} is reserved for mean ratings and 95\% confidence intervals. Ratings have not yet been collected; the conclusions below are based on automatic evaluation.
% 

%% file: table/03_encoder_comparison.tex
\begin{table}[!htbp]
\centering\small
\begin{tabular}{lr}
\toprule
Frontend recipe & COMET$\uparrow$ \\
\midrule
Earlier block-causal recipe & 0.689 \\
640\,ms chunkwise recipe & \textbf{0.856} \\
\bottomrule
\end{tabular}
\caption{S2TT-stage frontend recipes on FLEURS ($\tau=0.5$). Training recipes differ.}
\label{tab:encoder-comparison}
\end{table}

%% file: table/04_policy_ablation.tex
\begin{table}[!htbp]
\centering\small
\setlength{\tabcolsep}{2pt}
\begin{tabular*}{\columnwidth}{@{\extracolsep{\fill}}lrrr@{}}
\toprule
Policy & COMET$\uparrow$ & ASR-COMET$\uparrow$ & StartOffset$\downarrow$ \\
\midrule
% \multicolumn{4}{l}{\textit{FLEURS}} \\
Fixed $6$ & 0.804 & 0.745 & \textbf{5.42} \\
Fixed $8$ & 0.829 & 0.764 & 6.67 \\
MLP & \textbf{0.856} & \textbf{0.785} & 6.39 \\
\bottomrule
\end{tabular*}
\caption{Commitment policies on FLEURS. Fixed $K$ writes every $K$ source blocks; MLP uses $\tau=0.5$. StartOffset is in seconds.}
\label{tab:policy-ablation}
\end{table}

%% file: table/07_dynamic_threshold.tex
\begin{table}[!htbp]
\centering\small
\setlength{\tabcolsep}{2pt}
\begin{tabular*}{\columnwidth}{@{\extracolsep{\fill}}lrrr@{}}
\toprule
Policy & ASR-COMET$\uparrow$ & StartOffset$\downarrow$ & ATD$\downarrow$ \\
\midrule
Fixed $0.1$ & 0.761 & 4.73 & \textbf{5.49} \\
Fixed $0.2$ & \textbf{0.773} & 5.09 & 5.73 \\
Dynamic & \textbf{0.773} & \textbf{4.70} & 5.74 \\
\bottomrule
\end{tabular*}
\caption{Threshold policies on FLEURS. Latencies are in seconds; ATD excludes computation.}
\label{tab:dynamic-threshold}
\end{table}

%% file: table/05_fusion_ablation.tex
\begin{table}[!htbp]
\centering\small
\begin{tabular}{lrr}
\toprule
Conditioning & ASR-COMET$\uparrow$ & WER$\downarrow$ \\
\midrule
Final layer & 0.765 & 4.35 \\
Eight layers & \textbf{0.773} & \textbf{3.86} \\
\bottomrule
\end{tabular}
\caption{Speech-conditioning ablation on FLEURS with the dynamic policy.}
\label{tab:fusion-ablation}
\end{table}

%% file: text/07conclusion.tex
\section{Conclusion}
We presented \method{}, a supervised streaming speech-to-speech translation model that learns when to commit from semantically aligned trajectories. A 640\,ms chunkwise frontend retains pretrained convolution alignment, and weighted Thinker states continuously condition the Talker. The learned WRITE threshold adjusts commitment independently of input-block duration, with playback adaptation using buffered speech to gather additional context. FLEURS and internal 9-pair results establish strong translation performance against the evaluated streaming baselines; RealSI demonstrates the quality--latency tradeoff from a single checkpoint. FLEURS ablations support learned commitments and multi-layer conditioning. Human evaluation highlights the 4B model's audio quality, alongside competitive translation quality and speaker similarity.
\label{sec:main-end}

%% file: text/08limitations.tex
\section*{Limitations}
Our evaluation covers English and Chinese; broader multilingual transfer and long continuous conversations remain to be assessed. FLEURS and RealSI provide public evaluations, while the internal 9-pair set is not publicly released. Exact-match data audits do not rule out near-duplicate audio or overlap inherited from pretraining. Automatic ASR, speaker-similarity, and naturalness estimators may have language-dependent biases.

%% file: text/10implementation.tex
\section{Training Data and Supervision}
\label{sec:implementation}
\subsection{Shared Speech-Pair Construction}
The main data pair real source recordings with synthetic target speech. Starting from ASR transcripts, Qwen3-Max produces target translations and CosyVoice3.5 synthesizes target audio. We construct the two translation directions from real recordings in their respective source languages, rather than reversing synthetic target speech into additional source examples. Offline adaptation uses these complete-utterance pairs to train the speech-generation components and conditioning interface.

For streaming supervision, Qwen3.7-Max additionally divides the source and target transcripts into aligned semantic units. Forced alignment maps the units to the two waveforms, and the pipeline checks transcript reconstruction, monotonic timestamps, valid spans, and agreement between recorded and actual durations. The semantic-alignment mixture permits an 80\,ms boundary tolerance. Source boundaries are rounded up to the 640\,ms observation grid, without an additional block delay.

\subsection{Semantic Segmentation and Validation}
\label{app:semantic-labeling}
Qwen3.7-Max annotates paired transcripts in two passes, with separate prompts for English-to-Chinese and Chinese-to-English. Both passes use temperature zero with thinking disabled. The indexed units preserve the original text: English words, numbers, and punctuation, and Chinese characters with contiguous Latin or numeric expressions kept intact. Spaces are retained so that concatenation recovers the input exactly; these units are distinct from Thinker subword tokens. The first pass receives indexed tokens and proposes paired, monotonically increasing segment endpoints. The prompt favors the earliest semantically supported translation spans rather than equal-length or punctuation-only splits. It preserves names, numbers, negation, and fixed expressions, and allows larger spans when local word-order differences prevent a shorter alignment.

\paragraph{Prefix-supported commitment.}
The central criterion is directional: a target segment may express only information supported by the source prefix available at that boundary. The prompt explicitly prohibits introducing a person, action, object, number, or judgment whose source evidence occurs later. When a target phrase would anticipate such information, the annotator must merge adjacent segments until the source supplies it. Thus, the same bilingual sentence pair can require different boundaries in the two translation directions.

\paragraph{Revision with exact text.}
The second pass receives both full transcripts and the draft segmentation and returns revised pairs of continuous original-text substrings. It again checks prefix support while encouraging short aligned units, typically 4--12 Chinese characters and 2--8 English words; these are prompt preferences rather than fixed segment lengths. The revision may adjust boundaries but may not paraphrase, omit, or insert text.

\paragraph{Validation and retries.}
Deterministic checks require paired nonempty segments, monotonic boundaries, and exact reconstruction of both transcripts by concatenation. Punctuation-only fragments are merged with neighboring segments. Missing or duplicate identifiers, reconstruction mismatches, and failed review or monotonicity checks are recorded for retry and validation. The annotator also records fallback and review flags, distinguishing a returned segmentation from a successfully revised one. Forced alignment and ASR-based acoustic filtering follow these transcript-level checks.

\subsection{From Semantic Units to Streaming Trajectories}
\label{app:trajectory-construction}
The following procedure summarizes transcript annotation and its conversion into streaming supervision. Text checks operate on the original strings; acoustic checks use forced alignment associated with the corresponding source and target waveforms. Failed records are retried or excluded before packing.

\begin{figure}[!t]
\noindent\begin{minipage}{\columnwidth}
\small
\begin{tabularx}{\columnwidth}{@{}r@{\hspace{0.6em}}X@{}}
\toprule
\multicolumn{2}{@{}l@{}}{\textbf{Procedure: Streaming data construction}}\\
\midrule
& \textbf{Input:} Speech pair $(x,\widetilde{x})$, transcripts $(u,v)$, direction $d$, and training stage.\\
& \textbf{Output:} A validated READ/WRITE trajectory, or rejection.\\
1 & Index lossless text units in $u$ and $v$.\\
2 & $D\gets\operatorname{QwenDraft}(u,v,d)$: propose paired monotonic endpoints with source-prefix support.\\
3 & $D\gets\operatorname{QwenRevise}(u,v,D,d)$: revise into exact original-text substrings.\\
4 & Merge punctuation-only mismatches; check paired nonempty spans, transcript reconstruction, and review flags. Retry or reject failures.\\
5 & Obtain source and target forced alignments; check text correspondence, timestamp order, and waveform bounds.\\
6 & Recover source- and target-unit endpoints from the alignments.\\
7 & Round source endpoints up to the 640\,ms grid.\\
8 & Cap boundaries at the utterance end and place the final boundary there.\\
9 & Merge units sharing a source-grid endpoint and concatenate their paired target spans.\\
10 & For speech adaptation, partition the target codec sequence at the merged target endpoints on its 12.5\,Hz timeline.\\
11 & \textbf{For each} merged unit: consume source blocks, label internal decision points READ, and label its endpoint WRITE.\\
12 & Attach the target text and, for speech adaptation, the corresponding codec span; flush residual source input at end-of-input.\\
13 & Apply stage-specific duration, coverage, and ASR-consistency checks; pack retained trajectories with independent sample boundaries.\\
\bottomrule
\end{tabularx}
\end{minipage}
\end{figure}

Quantization can map several semantic boundaries to one observation point. Merging their paired spans avoids zero-duration source turns and preserves transcript order. Target codec spans are recomputed after this merge and cover the target sequence without gaps or overlaps. At inference, the learned gate selects commitments. Within a turn, target text and codec labels supervise generation, while READ/WRITE targets supervise the auxiliary gate rather than adding language-model tokens.

\subsection{Stage-Specific Mixtures}
\paragraph{Chunkwise encoder adaptation.}
The first streaming S2TT substage reuses the original utterance-level S2TT training pool of approximately 3M hours. Semantic segmentation and READ/WRITE labels are introduced only in the subsequent commitment-learning substage.

\paragraph{S2TT commitment learning.}
The semantic-trajectory mixture contains approximately 17k source-speech hours after upsampling. It combines the main bilingual corpus with approximately 20 hours of separate 9-pair training speech and 10 hours of retained FLEURS training speech; both supplements are upsampled tenfold and interleaved with the main corpus.

\paragraph{Offline speech adaptation.}
Offline S2ST uses approximately 300k hours of source speech in both training segments. After 5,000 updates, continuation expands this pool with supplemental FLEURS training speech upsampled tenfold. The continuation pool also includes earlier training and validation records; those earlier validation records are distinct from the held-out policy-selection data.

\paragraph{Streaming speech adaptation.}
The final pool contains approximately 13k hours of real source speech, with roughly 6.5k hours per direction. Direction balancing uses source duration rather than equal utterance counts. We supplement this pool with approximately 10 hours of retained FLEURS training speech upsampled tenfold. Both Talker adaptation and joint streaming training use this mixture.

The 9-pair training and test sets are split by textual content but are not strictly speaker-disjoint. We retain the text-disjoint training split for S2TT adaptation, but exclude it from both offline and streaming S2ST training to avoid inflating speaker similarity (SIM) through speaker overlap.

Streaming filtering uses source/target ASR-error cutoffs of 0.32/0.16 for Chinese-to-English and 0.4375/0.13125 for English-to-Chinese; it retains fewer FLEURS examples than the offline mixture.

\subsection{Data Separation and Policy Selection}
The FLEURS audit covers all 5,766 supplemental source records before filtering and packing; their paths belong to the training split, and their audio basenames and normalized text have no exact overlap with evaluation. We additionally scan the packed training manifest of the streaming text initialization: no exact canonical source path, audio basename, or normalized full target matches the 9-pair, FLEURS, or validation manifests. Both offline continuation and streaming S2ST retain FLEURS training data. Policy screening uses 128 official FLEURS validation recordings and 458 sentence-disjoint recordings for confirmation. Fixed-schedule controls use only validation onset for selecting block intervals; test quality is not used for selection.

\section{Training Stages and Optimization}
\label{app:training}
Table~\ref{tab:training-stages} groups 4B adaptation into streaming S2TT, offline S2ST, and streaming S2ST. S2TT comprises encoder adaptation and commitment learning; streaming S2ST comprises Talker and E2E substages. Offline S2ST retains one objective across its initial and continuation segments. The speech interface comprises fusion, projection, and codec embeddings/output head. Only the modules listed for each substage are updated. The Thinker has 32 layers and width 2,560. Flow matching and the vocoder remain frozen throughout adaptation.

\input{table/08_training_stages}

\subsection{Pretrained Initialization}
\label{app:initialization}
The initialization model combines a Qwen3.5-4B\footnote{\url{https://huggingface.co/Qwen/Qwen3.5-4B}} Thinker with a speech encoder initialized from SenseVoice\footnote{\url{https://github.com/QwenAudio/SenseVoice}} and is pretrained on approximately 3M hours of S2TT data. The Talker is initialized from CosyVoice\footnote{\url{https://github.com/QwenAudio/CosyVoice}} and pretrained on approximately 300k hours of S2ST data. Speech pretraining consists of 36,000 Talker-training updates with a learning rate of $10^{-4}$, followed by 1,600 end-to-end updates with a learning rate of $5\times10^{-6}$. These pretrained components provide the starting point for the adaptation stages described below.

\subsection{Streaming S2TT Adaptation}
\paragraph{Encoder adaptation.}
We first adapt the chunkwise encoder and adaptor together with Thinker LoRA using utterance-level S2TT supervision, without interleaved turns or READ/WRITE gate training. This substage reuses the original S2TT pool of approximately 3M hours. It uses an effective batch of 128 packed sequences, tensor parallelism one, sequence length 8,192, and 1,000 warmup updates. The Thinker LoRA learning rate is $10^{-5}$; encoder and adaptor learning rates are $2\times10^{-5}$. LoRA has rank 32 and scaling parameter 64. We use the 10,000-update checkpoint and merge its LoRA weights into the Thinker before commitment learning.

\paragraph{Commitment learning.}
From the encoder-adapted model, we train the full Thinker and READ/WRITE gate for 10,000 updates while freezing the chunkwise encoder and adaptor. This substage uses the semantic-trajectory mixture in Appendix~\ref{sec:implementation}, an effective batch of eight packed sequences, tensor parallelism two, and 1,000 warmup updates. The Thinker learning rate is $5\times10^{-6}$ and the gate learning rate is $10^{-4}$. Gate-to-Thinker gradients are scaled by 0.1. The resulting checkpoint supplies the frontend, Thinker, and gate for streaming S2ST.

\subsection{Offline S2ST Adaptation}
The source frontend and Thinker remain frozen while the Talker, codec embeddings/output head, projection, and fusion are trained. This is one speech-adaptation stage with 15,000 updates along the reported checkpoint lineage: 5,000 initial updates followed by 10,000 continuation updates. The continuation expands the paired-speech pool with supplemental training records and starts a fresh optimizer, scheduler, and data traversal; its objective and trainable modules are unchanged. The final offline checkpoint also supplies the speech components for streaming S2ST.

Both segments use an effective batch of 64, a Talker learning rate of $5\times10^{-6}$, projection/fusion learning rates of $10^{-4}$, and 100 warmup updates. Their cosine learning-rate schedules span 10,000 and 68,000 updates, respectively; the reported model uses 5,000 and 10,000 updates from these segments. Both segments use approximately 300k hours of source speech.

\subsection{Streaming S2ST Adaptation}
\paragraph{Talker substage.}
We combine the frontend, Thinker, and gate from streaming S2TT with the Talker and speech interface from offline S2ST. We train the Talker, codec embeddings/output head, projection, and fusion for 5,000 updates while keeping the translation components frozen. Training uses the filtered streaming mixture, an effective batch of 64, a learning rate of $5\times10^{-6}$, and 1,000 warmup updates.

\paragraph{E2E substage.}
From the Talker checkpoint, we jointly update the encoder, adaptor, Thinker, gate, fusion, projection, and Talker with the text, codec, and gate loss defined in the main text. This substage uses a fresh optimizer without warmup, an effective batch of 64, sequence length 8,192. The main and gate learning rates are $10^{-6}$ and $10^{-5}$, respectively. The run reaches 5,000 updates; the reported streaming checkpoint is taken at 4,000. Both substages use the same filtered streaming mixture, seed 1024, and BF16 model computation. E2E here denotes source-to-codec optimization; flow matching and the vocoder stay frozen.

\section{Model and Trajectory Details}
\label{app:model}
\subsection{Chunkwise Frontend}
The streaming encoder recomputes growing source prefixes, so computation increases with recording length. Its 31-tap FSMN convolutions retain centered tap alignment while masking unavailable right context. The adaptor uses eight-position causal blocks: a closed block cannot attend to future adaptor states. First-layer encoder lookahead can propagate within a block, but later layers cannot access the next block's contextualized states.

Encoder states are spaced at 40\,ms. For block size $B=16$, lookahead $R=4$, and zero-based position $i$, let $e(i)=B(\lfloor i/B\rfloor+1)-1$. Layer one permits keys $j\leq e(i)+R$; later layers permit $j\leq e(i)$. Fixed log-Mel normalization avoids a recording-wide maximum. Packed examples have independent attention and convolution boundaries.

The waveform frontend uses a 16\,kHz sampling rate, a 400-sample (25\,ms) Hann window, and a 160-sample (10\,ms) hop. Centered STFT framing (\texttt{center=True}) pads the waveform by 200 samples on each side using reflection. Frame $k$ is centered at sample $160k$ and covers samples $160k-200$ through $160k+199$. Following Whisper, we discard the final STFT frame before computing log-Mel features.

Let $b$ denote a nominal prefix endpoint after adding the encoder's 160\,ms lookahead, aligned to the 10\,ms frame grid. The last retained frame is centered at $b-160$ samples, so its analysis window extends through sample $b+39$. It therefore requires 40 additional waveform samples beyond $b$:
\[
\Delta_{\mathrm{STFT}}=\frac{400/2-160}{16{,}000}\,\mathrm{s}=2.5\,\mathrm{ms}.
\]
These samples prevent reflection at a temporary prefix boundary from replacing real right-context audio in the last retained frame. At the actual end of the utterance, the frontend uses its boundary padding without waiting for further input.

The total waveform lookahead is thus 162.5\,ms beyond each non-final chunk endpoint; the first complete 640\,ms block requires 802.5\,ms of received source audio. Only completed-block states are emitted, and this frontend waiting time is included in computation-aware StartOffset.

\subsection{Gate and Layer Fusion}
The gate takes $h_t\in\mathbb{R}^{2560}$ and computes
\begin{align}
u_t&=\operatorname{Dropout}(\operatorname{SiLU}(W_1\operatorname{LN}(h_t)+b_1)),\\
p_t&=\operatorname{softmax}(W_2u_t+b_2)_{\mathrm{WRITE}}.
\end{align}
Its intermediate width is 256, dropout is 0.1, and total size is approximately 0.66 million parameters.

Fusion uses signed, channelwise coefficients without softmax normalization or input-dependent routing. It combines intermediate post-block residual states with the final normalized layer-32 state. Products and accumulation use FP32 before conversion to the hidden-state dtype. Training captures autograd-connected states, including during activation recomputation; inference follows the same state convention. Setting the first seven layer vectors to zero and the final vector to one recovers final-state conditioning.

These dimensions and layer selections describe the 4B model used in the component ablations. The 35B-total, 3B-active MoE variants in Tables~\ref{tab:s2st-offline} and~\ref{tab:s2st} have 40 Thinker layers of width 2,048 and fuse states from layers 8, 16, 24, 32, and 40.

\subsection{Streaming Trajectories}
Each WRITE opens an assistant turn, terminated by the existing end-of-turn token; the next source span opens a user/audio turn. READ/WRITE labels do not add language-model tokens. Codec supervision uses five delay positions, a beginning token, codec targets, and an ending token for each WRITE. The codec recipe inserts neither a pause token nor a second beginning token, and permits a final conditioning drain.

\section{Inference and Evaluation}
\label{app:evaluation}
\subsection{Rendering and Scoring}
The default streaming renderer uses eight-codec flow-matching chunks, three-codec flow-matching lookahead, eight-mel vocoder lookahead, and a one-codec EOS hold.

For FLEURS, we prepare a shared set of waveforms before replay using one gain per utterance. The gain raises RMS toward 0.0743, subject to an amplification peak limit of 0.95, without attenuating the input. RMS and peak are computed over the complete waveform; sample rates and durations are unchanged. All systems use these saved inputs. X-Translator retains its existing per-utterance input-gain stage, capped at peak 0.99.

Local ASR uses Whisper-large-v3 for English and SeACo-Paraformer-large for Chinese; generated speech is resampled to 16\,kHz before Chinese ASR. Empty outputs remain in translation scoring; unavailable speech observations and onsets are excluded from their corresponding averages. The main tables use the same dataset manifests and scoring pipeline across systems. The 35B/3B offline variant uses its archived legacy speaker-conditioning configuration. One English-to-Chinese example has no usable audio; it remains in text and ASR-based scoring with an empty ASR transcript, while SIM and UTMOS use the 995 available waveforms (996 for Chinese-to-English).

For the 35B/3B fixed streaming model on 9-pair, one example has no audio; SIM and UTMOS use the 1,991 available waveforms.

\paragraph{Corpus-level synthesis fidelity.}
WER/CER uses the model's generated translation as the reference and the ASR transcription of its output speech as the hypothesis. Within each direction, we pool substitution, deletion, and insertion counts over the evaluation set:
\[
E_d=100\,\frac{\sum_i(S_i+D_i+I_i)}{\sum_i N_i},
\]
where $N_i$ is the reference length. English uses word units and Chinese uses character units. The reported WER column is the unweighted mean of English WER and Chinese CER; it is not an average of per-utterance error rates.

\subsection{Evaluation Data and Latency}
FLEURS quality results use the public English--Chinese test data described in Section~\ref{sec:setup}. The internal 9-pair evaluation uses 996 examples per direction, shared across systems. The added 9-pair training sources used in text adaptation are distinct from these evaluation manifests, subject to the audit scope described above. RealSI quality and latency use the public sentence-level evaluation data, with direction-level scores averaged equally. Public FLEURS and RealSI results provide the reproducible evaluation alongside the internal test set.

Computation-aware StartOffset begins at the start of real-time source replay and ends when the first nonempty waveform packet is emitted. It includes waiting for source audio, source-speaker conditioning, Thinker and Talker computation, flow matching, and vocoding. Model loading, warmup, networking, and device playback buffering are excluded. We use raw packet emission without forced-alignment correction to the first packet, and concatenate the same emitted waveforms for quality scoring.

\subsection{Baseline Configurations}
We evaluate UniSS in its Quality and Performance modes and SimulS2ST-Omni at its m2 setting for the main tables, using 1,000\,ms source segments and a latency multiplier of two. The RealSI curve additionally sweeps latency multipliers m1--m5. The IWSLT VAD implementation is pinned to its 2025 release; its English-to-Chinese recipe is additionally adapted to Chinese-to-English. SeamlessStreaming uses 320\,ms source segments, decision threshold 0.5, FP32 inference, and no silence stripping. SimulS2ST-Omni uses deterministic decoding with four Thinker beams and maximum text/codec generation lengths of 256/500 tokens, retaining the official minimum final-chunk duration of 0.32\,s. The IWSLT (VAD) run uses 250\,ms input segments, a 1\,s step, voice threshold 0.1, and maximum unvoiced length 0.1\,s. Agent caches and Silero recurrent state are reset between utterances. Full commands, implementation revisions, and output traces are retained for reproducibility. UniSS is included in the quality comparison only.

\subsection{Parameter-Count Scope}
Tables~\ref{tab:s2st-offline} and~\ref{tab:s2st} report nominal language-backbone scales for LLM-based systems: UniSS uses a 1.5B backbone \citep{cheng2025uniss}, S3-S2ST a 4B backbone, X-Translator an 8B translation backbone, SimulS2ST-Omni a 3B Thinker, and Qwen3-Omni a 30B MoE Thinker with approximately 3B active parameters per token \citep{xu2025qwen3omni}. For these backbone entries, speech encoders, separate Talkers, speech tokenizers, waveform decoders, and other auxiliary modules are excluded. Seamless systems have no directly corresponding LLM backbone: dagger-marked entries therefore use their official published model-level sizes, 2.3B for SeamlessM4T-v2-large and 2.5B for SeamlessStreaming \citep{seamless2023}. IWSLT (VAD) uses the 2.3B SeamlessM4T-v2-large checkpoint for S2TT; this figure excludes its external VAD.

\section{Ablation Protocols}
\label{app:ablations}
\subsection{Frontend Recipes}
All main ablations use the same FLEURS test manifests. The historical encoder comparison reuses the earlier block-causal text model after 12,000 joint updates, initialized with a separately adapted encoder and an earlier streaming translation model. Its archived inference uses the original convolution padding, block-causal encoder attention, the original adaptor, and per-prefix log-mel clipping. We restore the archived inference sources and use 640\,ms input blocks, a fixed WRITE threshold of 0.5, greedy text decoding, and a maximum of 256 tokens per WRITE. The chunkwise comparator is the existing 10,000-update text model used to initialize our streaming system. The recipes use different initialization sources, frontend training, and data schedules. Text COMET is averaged equally over 646 English-to-Chinese and 945 Chinese-to-English examples. The historical run uses official FLEURS recordings matched to the manifest filenames, transcriptions, and durations.

\subsection{Commitment Policy}
Policy ablations use separate inference runs with speaker conditioning fixed from the initial source prefix; main-table runs refresh it at each WRITE.
All policies retain the same gate and cache computations, with only the READ/WRITE decisions changed. We use 10,000 paired sentence-cluster bootstrap draws, keeping recordings and directions of the same sentence together. For MLP minus fixed $K=8$, the 95\% ASR-COMET interval on FLEURS is $[0.014,0.029]$. One MLP example does not reach the final codec EOS; its generated text and emitted speech are included in scoring.

\subsection{Speech Conditioning}
The final-layer fusion control follows the final streaming Talker and joint-training stages. It uses playback-adaptive decoding, FM8, and source-speaker conditioning refreshed at each WRITE, matching the completed eight-layer branch.
Both fusion branches are evaluated on all 1,591 FLEURS examples (646 English-to-Chinese and 945 Chinese-to-English). We compute ASR-COMET over all utterances and WER/CER from pooled error and reference-token counts within each direction, then average the two direction-level scores. Table~\ref{tab:fusion-ablation} compares quality only, not runtime.

\subsection{Playback-Adaptive Threshold}
The FLEURS comparison in Table~\ref{tab:dynamic-threshold} uses 945 Chinese-to-English and 646 English-to-Chinese examples per policy, with direction-level scores averaged equally. We compute computation-unaware ATD from packet-level source-consumption positions and output durations using SimulEval's 300\,ms speech units \citep{ma-etal-2020-simuleval}. Adaptive and fixed $\tau=0.1$ decoding choose identical first-WRITE input positions; their small StartOffset difference reflects runtime variation. For adaptive minus fixed $\tau=0.1$, 5,000 paired utterance-bootstrap draws within each direction give an ASR-COMET gain of 0.013 with a 95\% interval of $[0.008,0.017]$.

\section{Human Evaluation}
\label{app:human}
We randomly sample 200 distinct FLEURS utterances per direction with paired outputs from \method{} (4B, fixed $\tau=0.5$), SeamlessStreaming, SimulS2ST-Omni (m2), X-Translator, and Hy-SimulMT (1.6\,s). Ten raters per direction each assess 20 disjoint utterances; each utterance and its five system outputs are judged by one rater. Source audio, reference translations, generated text, and translated audio are provided. Translation quality, audio quality, and source-speaker similarity are rated separately from 1 to 5, with higher scores indicating better quality.

We report the equally weighted mean of the two direction-specific scores with 95\% confidence intervals. To account for dependence among judgments from the same rater \citep{camp2023mos}, we use 20,000 rater-cluster bootstrap draws with seed 20261118. Each draw independently resamples ten raters with replacement within each direction, retains all their paired item and system ratings, and recomputes the bidirectional mean. The reported interval is the mean $\pm 1.96$ times the standard deviation of these bootstrap means. A score of zero denotes \emph{unable to judge} and is excluded only from that dimension. Translation and audio quality have 200 valid scores per system per direction. Speaker similarity excludes two scores for \method{} and one for X-Translator in English-to-Chinese, and one for SeamlessStreaming in Chinese-to-English.

%% file: table/08_training_stages.tex
\begin{table*}[!t]
\centering\footnotesize
\setlength{\tabcolsep}{3pt}
\renewcommand{\arraystretch}{1.15}
\begin{tabularx}{\textwidth}{@{}>{\hsize=1.15\hsize\linewidth=\hsize\raggedright\arraybackslash}X>{\hsize=.95\hsize\linewidth=\hsize\raggedright\arraybackslash}X>{\hsize=.95\hsize\linewidth=\hsize\raggedright\arraybackslash}Xcr c>{\hsize=.95\hsize\linewidth=\hsize\raggedright\arraybackslash}X@{}}
\toprule
Substage & Initialization & Updated modules & Training data & Updates & Batch size & Learning rate \\
\midrule
\multicolumn{7}{@{}l}{\textbf{I. Streaming S2TT adaptation}} \\
Encoder adaptation & Pretrained S2TT components & Encoder, adaptor, Thinker LoRA & $\sim$3M h & 10k & 128 & LoRA: $10^{-5}$; encoder/adaptor: $2\times10^{-5}$ \\
Commitment learning & Encoder-adapted model with LoRA merged & Thinker, READ/WRITE gate & $\sim$17k h & 10k & 8 & Thinker: $5\times10^{-6}$; gate: $10^{-4}$ \\
\midrule
\multicolumn{7}{@{}l}{\textbf{II. Offline S2ST adaptation}} \\
Talker adaptation & Pretrained source frontend, Thinker, and Talker; new speech interface & Talker, speech interface & $\sim$300k h & 15k & 64 & Talker: $5\times10^{-6}$; projection/fusion: $10^{-4}$ \\
\midrule
\multicolumn{7}{@{}l}{\textbf{III. Streaming S2ST adaptation}} \\
Talker & Frontend, Thinker, and gate from I; Talker and interface from II & Talker, speech interface & $\sim$13k h & 5k & 64 & $5\times10^{-6}$ \\
E2E & Streaming Talker checkpoint & Encoder, adaptor, Thinker, gate, Talker, speech interface & $\sim$13k h & 4k & 64 & main: $10^{-6}$; gate: $10^{-5}$ \\
\bottomrule
\end{tabularx}
\caption{Adaptation stages used to obtain the reported models.}
\label{tab:training-stages}
\vspace{2pt}
\begin{minipage}{\textwidth}
% \footnotesize
% Data sizes are source-speech hours including upsampling; batch sizes count packed sequences.
\end{minipage}
\end{table*}

%% file: references.bib
@inproceedings{ma-etal-2019-stacl,
    title = "{STACL}: Simultaneous Translation with Implicit Anticipation and Controllable Latency using Prefix-to-Prefix Framework",
    author = "Ma, Mingbo  and
      Huang, Liang  and
      Xiong, Hao  and
      Zheng, Renjie  and
      Liu, Kaibo  and
      Zheng, Baigong  and
      Zhang, Chuanqiang  and
      He, Zhongjun  and
      Liu, Hairong  and
      Li, Xing  and
      Wu, Hua  and
      Wang, Haifeng",
    editor = "Korhonen, Anna  and
      Traum, David  and
      M{\`a}rquez, Llu{\'i}s",
    booktitle = "Proceedings of the 57th Annual Meeting of the {A}ssociation for {C}omputational {L}inguistics",
    month = jul,
    year = "2019",
    address = "Florence, Italy",
    publisher = "Association for Computational Linguistics",
    url = "https://aclanthology.org/P19-1289/",
    doi = "10.18653/v1/P19-1289",
    pages = "3025--3036"
}

@inproceedings{zhang-etal-2022-learning,
    title = "Learning Adaptive Segmentation Policy for End-to-End Simultaneous Translation",
    author = "Zhang, Ruiqing  and
      He, Zhongjun  and
      Wu, Hua  and
      Wang, Haifeng",
    editor = "Muresan, Smaranda  and
      Nakov, Preslav  and
      Villavicencio, Aline",
    booktitle = "Proceedings of the 60th Annual Meeting of the {A}ssociation for {C}omputational {L}inguistics (Volume 1: Long Papers)",
    month = may,
    year = "2022",
    address = "Dublin, Ireland",
    publisher = "Association for Computational Linguistics",
    url = "https://aclanthology.org/2022.acl-long.542/",
    doi = "10.18653/v1/2022.acl-long.542",
    pages = "7862--7874"
}

@inproceedings{peters-etal-2018-deep,
    title = "Deep Contextualized Word Representations",
    author = "Peters, Matthew E.  and
      Neumann, Mark  and
      Iyyer, Mohit  and
      Gardner, Matt  and
      Clark, Christopher  and
      Lee, Kenton  and
      Zettlemoyer, Luke",
    editor = "Walker, Marilyn  and
      Ji, Heng  and
      Stent, Amanda",
    booktitle = "Proceedings of the 2018 Conference of the {N}orth {A}merican Chapter of the {A}ssociation for {C}omputational {L}inguistics: Human Language Technologies, Volume 1 (Long Papers)",
    month = jun,
    year = "2018",
    address = "New Orleans, Louisiana",
    publisher = "Association for Computational Linguistics",
    url = "https://aclanthology.org/N18-1202/",
    doi = "10.18653/v1/N18-1202",
    pages = "2227--2237"
}

@article{guerreiro-etal-2024-xcomet,
    title = "x{COMET}: Transparent Machine Translation Evaluation through Fine-grained Error Detection",
    author = "Guerreiro, Nuno M.  and
      Rei, Ricardo  and
      van Stigt, Daan  and
      Coheur, Luisa  and
      Colombo, Pierre  and
      Martins, Andr{\'e} F. T.",
    journal = "Transactions of the Association for Computational Linguistics",
    volume = "12",
    year = "2024",
    address = "Cambridge, MA",
    publisher = "MIT Press",
    url = "https://aclanthology.org/2024.tacl-1.54/",
    doi = "10.1162/tacl_a_00683",
    pages = "979--995"
}

@inproceedings{ma-etal-2020-simuleval,
    title = "{SIMULEVAL}: An Evaluation Toolkit for Simultaneous Translation",
    author = "Ma, Xutai  and
      Dousti, Mohammad Javad  and
      Wang, Changhan  and
      Gu, Jiatao  and
      Pino, Juan",
    editor = "Liu, Qun  and
      Schlangen, David",
    booktitle = "Proceedings of the 2020 Conference on Empirical Methods in Natural Language Processing: System Demonstrations",
    month = oct,
    year = "2020",
    address = "Online",
    publisher = "Association for Computational Linguistics",
    url = "https://aclanthology.org/2020.emnlp-demos.19/",
    doi = "10.18653/v1/2020.emnlp-demos.19",
    pages = "144--150"
}

@misc{seamless2023,
 title = {{Seamless}: Multilingual Expressive and Streaming Speech Translation},
 author = {{Seamless Communication} and Barrault, Loïc and Chung, Yu-An and Meglioli, Mariano Coria and Dale, David and Dong, Ning and Duppenthaler, Mark and Duquenne, Paul-Ambroise and Ellis, Brian and Elsahar, Hady and Haaheim, Justin and Hoffman, John and Hwang, Min-Jae and Inaguma, Hirofumi and Klaiber, Christopher and Kulikov, Ilia and Li, Pengwei and Licht, Daniel and Maillard, Jean and Mavlyutov, Ruslan and Rakotoarison, Alice and Sadagopan, Kaushik Ram and Ramakrishnan, Abinesh and Tran, Tuan and Wenzek, Guillaume and Yang, Yilin and Ye, Ethan and Evtimov, Ivan and Fernandez, Pierre and Gao, Cynthia and Hansanti, Prangthip and Kalbassi, Elahe and Kallet, Amanda and Kozhevnikov, Artyom and Gonzalez, Gabriel Mejia and Roman, Robin San and Touret, Christophe and Wong, Corinne and Wood, Carleigh and Yu, Bokai and Andrews, Pierre and Balioglu, Can and Chen, Peng-Jen and Costa-jussà, Marta R. and Elbayad, Maha and Gong, Hongyu and Guzmán, Francisco and Heffernan, Kevin and Jain, Somya and Kao, Justine and Lee, Ann and Ma, Xutai and Mourachko, Alex and Peloquin, Benjamin and Pino, Juan and Popuri, Sravya and Ropers, Christophe and Saleem, Safiyyah and Schwenk, Holger and Sun, Anna and Tomasello, Paden and Wang, Changhan and Wang, Jeff and Wang, Skyler and Williamson, Mary},
 year={2023}, eprint={2312.05187}, archivePrefix={arXiv}, url={https://arxiv.org/abs/2312.05187}}

@inproceedings{conneau2022fleurs, title = {{FLEURS}: Few-shot Learning Evaluation of Universal Representations of Speech}, url = {https://doi.org/10.1109/SLT54892.2023.10023141}, doi = {10.1109/SLT54892.2023.10023141}, booktitle={2022 IEEE Spoken Language Technology Workshop (SLT)}, publisher={IEEE}, author={Conneau, Alexis and Ma, Min and Khanuja, Simran and Zhang, Yu and Axelrod, Vera and Dalmia, Siddharth and Riesa, Jason and Rivera, Clara and Bapna, Ankur}, year={2023}, month=Jan, pages = {798--805} }

@misc{xu2025omni,
 title = {{Qwen2.5-Omni} Technical Report},
 author = {Jin Xu and Zhifang Guo and Jinzheng He and Hangrui Hu and Ting He and Shuai Bai and Keqin Chen and Jialin Wang and Yang Fan and Kai Dang and Bin Zhang and Xiong Wang and Yunfei Chu and Junyang Lin},
 year={2025}, eprint={2503.20215}, archivePrefix={arXiv}, url={https://arxiv.org/abs/2503.20215}}

@inproceedings{saeki2022utmos,
  title     = {{UTMOS: UTokyo-SaruLab System for VoiceMOS Challenge 2022}},
  author    = {Takaaki Saeki and Detai Xin and Wataru Nakata and Tomoki Koriyama and Shinnosuke Takamichi and Hiroshi Saruwatari},
  year      = {2022},
  booktitle = {{Interspeech 2022}},
  pages     = {4521--4525},
  doi       = {10.21437/Interspeech.2022-439},
  issn      = {2958-1796},
  url = {https://www.isca-archive.org/interspeech_2022/saeki22c_interspeech.html}
}

@misc{ma2023emma,
 title={Efficient Monotonic Multihead Attention},
 author={Ma, Xutai and Sun, Anna and Ouyang, Siqi and Inaguma, Hirofumi and Tomasello, Paden},
 year={2023}, eprint={2312.04515}, archivePrefix={arXiv}, url={https://arxiv.org/abs/2312.04515}}

@inproceedings{he2026simuls2st,
  title = {{SimulS2ST-Omni}: Data-Efficient Streaming Speech-to-Speech Translation via Explicit Trajectory Supervision},
  author = {He, Rongshen and Liang, Xinyu and Chen, Dekun and Li, Jiaqi and Chen, Mingjie and Wu, Zhizheng},
  booktitle = {Proceedings of the 2026 Conference on Empirical Methods in Natural Language Processing},
  year = {2026},
  url = {https://aclanthology.org/2026.emnlp-main.1135/}
}

@inproceedings{cheng2025uniss,
 author={Cheng, Sitong and Bian, Weizhen and Wang, Xinsheng and Yuan, Ruibin and Chen, Jianyi and Yin, Shunshun and Guo, Yike and Xue, Wei},
 title={{UniSS}: Unified Expressive Speech-to-Speech Translation with Your Voice}, year = {2026}, url = {https://openreview.net/forum?id=5o0ZvYzh6B},
  booktitle = {The Fourteenth International Conference on Learning Representations}
}

@misc{zhao2026xtranslator,
 author={Zhao, Yuxiang and Zhang, Yichi and An, Yanjie and Zhu, Yanqiao and Liu, Zhanxun and Chen, Yushen and Zheng, Qixi and Zhu, Haina and Xiao, Yunchong and Deng, Keqi and Fan, Shuai and Yu, Kai and Chen, Xie},
 title={{X-Translator}: A Real-Time Multilingual Speaker-Aware Speech-to-Speech Translation System}, year={2026}, eprint={2607.17544}, archivePrefix={arXiv}, url={https://arxiv.org/abs/2607.17544},
  primaryClass = {eess.AS}
}

@misc{xu2025qwen3omni,
 author = {Jin Xu and Zhifang Guo and Hangrui Hu and Yunfei Chu and Xiong Wang and Jinzheng He and Yuxuan Wang and Xian Shi and Ting He and Xinfa Zhu and Yuanjun Lv and Yongqi Wang and Dake Guo and He Wang and Linhan Ma and Pei Zhang and Xinyu Zhang and Hongkun Hao and Zishan Guo and Baosong Yang and Bin Zhang and Ziyang Ma and Xipin Wei and Shuai Bai and Keqin Chen and Xuejing Liu and Peng Wang and Mingkun Yang and Dayiheng Liu and Xingzhang Ren and Bo Zheng and Rui Men and Fan Zhou and Bowen Yu and Jianxin Yang and Le Yu and Jingren Zhou and Junyang Lin},
 title = {{Qwen3-Omni} Technical Report}, year={2025}, eprint={2509.17765}, archivePrefix={arXiv}, url={https://arxiv.org/abs/2509.17765}}

@inproceedings{lee-etal-2022-direct,
    title = "Direct Speech-to-Speech Translation With Discrete Units",
    author = "Lee, Ann  and
      Chen, Peng-Jen  and
      Wang, Changhan  and
      Gu, Jiatao  and
      Popuri, Sravya  and
      Ma, Xutai  and
      Polyak, Adam  and
      Adi, Yossi  and
      He, Qing  and
      Tang, Yun  and
      Pino, Juan  and
      Hsu, Wei-Ning",
    editor = "Muresan, Smaranda  and
      Nakov, Preslav  and
      Villavicencio, Aline",
    booktitle = "Proceedings of the 60th Annual Meeting of the {A}ssociation for {C}omputational {L}inguistics (Volume 1: Long Papers)",
    month = may,
    year = "2022",
    address = "Dublin, Ireland",
    publisher = "Association for Computational Linguistics",
    url = "https://aclanthology.org/2022.acl-long.235/",
    doi = "10.18653/v1/2022.acl-long.235",
    pages = "3327--3339"
}

@inproceedings{inaguma-etal-2023-unity,
    title = "{U}nit{Y}: Two-pass Direct Speech-to-speech Translation with Discrete Units",
    author = "Inaguma, Hirofumi  and
      Popuri, Sravya  and
      Kulikov, Ilia  and
      Chen, Peng-Jen  and
      Wang, Changhan  and
      Chung, Yu-An  and
      Tang, Yun  and
      Lee, Ann  and
      Watanabe, Shinji  and
      Pino, Juan",
    editor = "Rogers, Anna  and
      Boyd-Graber, Jordan  and
      Okazaki, Naoaki",
    booktitle = "Proceedings of the 61st Annual Meeting of the {A}ssociation for {C}omputational {L}inguistics (Volume 1: Long Papers)",
    month = jul,
    year = "2023",
    address = "Toronto, Canada",
    publisher = "Association for Computational Linguistics",
    url = "https://aclanthology.org/2023.acl-long.872/",
    doi = "10.18653/v1/2023.acl-long.872",
    pages = "15655--15680"
}

@inproceedings{zhang-etal-2024-streamspeech,
    title = "{S}tream{S}peech: Simultaneous Speech-to-Speech Translation with Multi-task Learning",
    author = "Zhang, Shaolei  and
      Fang, Qingkai  and
      Guo, Shoutao  and
      Ma, Zhengrui  and
      Zhang, Min  and
      Feng, Yang",
    editor = "Ku, Lun-Wei  and
      Martins, Andre  and
      Srikumar, Vivek",
    booktitle = "Proceedings of the 62nd Annual Meeting of the {A}ssociation for {C}omputational {L}inguistics (Volume 1: Long Papers)",
    month = aug,
    year = "2024",
    address = "Bangkok, Thailand",
    publisher = "Association for Computational Linguistics",
    url = "https://aclanthology.org/2024.acl-long.485/",
    doi = "10.18653/v1/2024.acl-long.485",
    pages = "8964--8986"
}

@inproceedings{deng-etal-2025-simuls2s,
    title = "{S}imul{S}2{S}-{LLM}: Unlocking Simultaneous Inference of Speech {LLM}s for Speech-to-Speech Translation",
    author = "Deng, Keqi  and
      Chen, Wenxi  and
      Chen, Xie  and
      Woodland, Phil",
    editor = "Che, Wanxiang  and
      Nabende, Joyce  and
      Shutova, Ekaterina  and
      Pilehvar, Mohammad Taher",
    booktitle = "Proceedings of the 63rd Annual Meeting of the {A}ssociation for {C}omputational {L}inguistics (Volume 1: Long Papers)",
    month = jul,
    year = "2025",
    address = "Vienna, Austria",
    publisher = "Association for Computational Linguistics",
    url = "https://aclanthology.org/2025.acl-long.817/",
    doi = "10.18653/v1/2025.acl-long.817",
    pages = "16718--16734",
    ISBN = "979-8-89176-251-0"
}

@inproceedings{zhang-feng-2023-end,
    title = "End-to-End Simultaneous Speech Translation with Differentiable Segmentation",
    author = "Zhang, Shaolei  and
      Feng, Yang",
    editor = "Rogers, Anna  and
      Boyd-Graber, Jordan  and
      Okazaki, Naoaki",
    booktitle = "Findings of the {A}ssociation for {C}omputational {L}inguistics: {ACL} 2023",
    month = jul,
    year = "2023",
    address = "Toronto, Canada",
    publisher = "Association for Computational Linguistics",
    url = "https://aclanthology.org/2023.findings-acl.485/",
    doi = "10.18653/v1/2023.findings-acl.485",
    pages = "7659--7680"
}

@article{papi-etal-2025-real,
    title = "How ``Real'' is Your Real-Time Simultaneous Speech-to-Text Translation System?",
    author = "Papi, Sara  and
      Pol{\'a}k, Peter  and
      Mach{\'a}{\v{c}}ek, Dominik  and
      Bojar, Ond{\v{r}}ej",
    journal = "Transactions of the Association for Computational Linguistics",
    volume = "13",
    year = "2025",
    address = "Cambridge, MA",
    publisher = "MIT Press",
    url = "https://aclanthology.org/2025.tacl-1.14/",
    doi = "10.1162/tacl_a_00740",
    pages = "281--313"
}

@inproceedings{xu-etal-2025-ca,
    title = "{CA}*: Addressing Evaluation Pitfalls in Computation-Aware Latency for Simultaneous Speech Translation",
    author = "Xu, Xi  and
      Xu, Wenda  and
      Ouyang, Siqi  and
      Li, Lei",
    editor = "Chiruzzo, Luis  and
      Ritter, Alan  and
      Wang, Lu",
    booktitle = "Findings of the {A}ssociation for {C}omputational {L}inguistics: {NAACL} 2025",
    month = apr,
    year = "2025",
    address = "Albuquerque, New Mexico",
    publisher = "Association for Computational Linguistics",
    url = "https://aclanthology.org/2025.findings-naacl.393/",
    doi = "10.18653/v1/2025.findings-naacl.393",
    pages = "7077--7082",
    ISBN = "979-8-89176-195-7"
}

@inproceedings{post-2018-call,
    title = "A Call for Clarity in Reporting {BLEU} Scores",
    author = "Post, Matt",
    editor = "Bojar, Ond{\v{r}}ej  and
      Chatterjee, Rajen  and
      Federmann, Christian  and
      Fishel, Mark  and
      Graham, Yvette  and
      Haddow, Barry  and
      Huck, Matthias  and
      Yepes, Antonio Jimeno  and
      Koehn, Philipp  and
      Monz, Christof  and
      Negri, Matteo  and
      N{\'e}v{\'e}ol, Aur{\'e}lie  and
      Neves, Mariana  and
      Post, Matt  and
      Specia, Lucia  and
      Turchi, Marco  and
      Verspoor, Karin",
    booktitle = "Proceedings of the Third Conference on Machine Translation: Research Papers",
    month = oct,
    year = "2018",
    address = "Brussels, Belgium",
    publisher = "Association for Computational Linguistics",
    url = "https://aclanthology.org/W18-6319/",
    doi = "10.18653/v1/W18-6319",
    pages = "186--191"
}

@inproceedings{rei-etal-2020-comet,
    title = "{COMET}: A Neural Framework for {MT} Evaluation",
    author = "Rei, Ricardo  and
      Stewart, Craig  and
      Farinha, Ana C  and
      Lavie, Alon",
    editor = "Webber, Bonnie  and
      Cohn, Trevor  and
      He, Yulan  and
      Liu, Yang",
    booktitle = "Proceedings of the 2020 Conference on Empirical Methods in Natural Language Processing ({EMNLP})",
    month = nov,
    year = "2020",
    address = "Online",
    publisher = "Association for Computational Linguistics",
    url = "https://aclanthology.org/2020.emnlp-main.213/",
    doi = "10.18653/v1/2020.emnlp-main.213",
    pages = "2685--2702"
}

@inproceedings{jia-etal-2022-cvss,
    title = "{CVSS} Corpus and Massively Multilingual Speech-to-Speech Translation",
    author = "Jia, Ye  and
      Tadmor Ramanovich, Michelle  and
      Wang, Quan  and
      Zen, Heiga",
    editor = "Calzolari, Nicoletta  and
      B{\'e}chet, Fr{\'e}d{\'e}ric  and
      Blache, Philippe  and
      Choukri, Khalid  and
      Cieri, Christopher  and
      Declerck, Thierry  and
      Goggi, Sara  and
      Isahara, Hitoshi  and
      Maegaard, Bente  and
      Mariani, Joseph  and
      Mazo, H{\'e}l{\`e}ne  and
      Odijk, Jan  and
      Piperidis, Stelios",
    booktitle = "Proceedings of the Thirteenth Language Resources and Evaluation Conference",
    month = jun,
    year = "2022",
    address = "Marseille, France",
    publisher = "European Language Resources Association",
    url = "https://aclanthology.org/2022.lrec-1.720/",
    pages = "6691--6703"
}

@inproceedings{agostinelli-etal-2025-findings,
    title = "Findings of the {IWSLT} 2025 Evaluation Campaign",
    author = {Abdulmumin, Idris  and
      Agostinelli, Victor  and
      Alum{\"a}e, Tanel  and
      Anastasopoulos, Antonios  and
      Bentivogli, Luisa  and
      Bojar, Ond{\v{r}}ej  and
      Borg, Claudia  and
      Bougares, Fethi  and
      Cattoni, Roldano  and
      Cettolo, Mauro  and
      Chen, Lizhong  and
      Chen, William  and
      Dabre, Raj  and
      Est{\`e}ve, Yannick  and
      Federico, Marcello  and
      Fishel, Mark  and
      Gaido, Marco  and
      Javorsk{\'y}, D{\'a}vid  and
      Kasztelnik, Marek  and
      Kponou, Fortun{\'e}  and
      Krubi{\'n}ski, Mateusz  and
      Kin Lam, Tsz  and
      Liu, Danni  and
      Matusov, Evgeny  and
      Kumar Maurya, Chandresh  and
      McCrae, John P.  and
      Mdhaffar, Salima  and
      Moslem, Yasmin  and
      Murray, Kenton  and
      Nakamura, Satoshi  and
      Negri, Matteo  and
      Niehues, Jan  and
      Kr. Ojha, Atul  and
      Ortega, John E.  and
      Papi, Sara  and
      Pecina, Pavel  and
      Pol{\'a}k, Peter  and
      Po{\l}e{\'c}, Piotr  and
      Sankar, Ashwin  and
      Savoldi, Beatrice  and
      Sethiya, Nivedita  and
      Sikasote, Claytone  and
      Sperber, Matthias  and
      St{\"u}ker, Sebastian  and
      Sudoh, Katsuhito  and
      Thompson, Brian  and
      Turchi, Marco  and
      Waibel, Alex  and
      Wilken, Patrick  and
      Zevallos, Rodolfo  and
      Zouhar, Vil{\'e}m  and
      Z{\"u}fle, Maike},
    editor = "Salesky, Elizabeth  and
      Federico, Marcello  and
      Anastasopoulos, Antonis",
    booktitle = "Proceedings of the 22nd International Conference on Spoken Language Translation ({IWSLT} 2025)",
    month = jul,
    year = "2025",
    address = "Vienna, Austria (in-person and online)",
    publisher = "Association for Computational Linguistics",
    url = "https://aclanthology.org/2025.iwslt-1.44/",
    doi = "10.18653/v1/2025.iwslt-1.44",
    pages = "412--481",
    ISBN = "979-8-89176-272-5"
}

@InProceedings{labiausse2025hibiki,
  title = 	 {High-Fidelity Simultaneous Speech-To-Speech Translation},
  author =       {Labiausse, Tom and Mazar\'{e}, Laurent and Grave, Edouard and D\'{e}fossez, Alexandre and Zeghidour, Neil},
  booktitle = 	 {Proceedings of the 42nd International Conference on Machine Learning},
  pages = 	 {32116--32129},
  year = 	 {2025},
  editor = 	 {Singh, Aarti and Fazel, Maryam and Hsu, Daniel and Lacoste-Julien, Simon and Berkenkamp, Felix and Maharaj, Tegan and Wagstaff, Kiri and Zhu, Jerry},
  volume = 	 {267},
  series = 	 {Proceedings of Machine Learning Research},
  month = 	 {13--19 Jul},
  publisher =    {PMLR},
  url = 	 {https://proceedings.mlr.press/v267/labiausse25a.html},
  }

@InProceedings{labiausse2026hibikizero,
  title = 	 {Simultaneous Speech-to-Speech Translation Without Aligned Data},
  author =       {Labiausse, Tom and Fabre, Romain and Est\`{e}ve, Yannick and D\'{e}fossez, Alexandre and Zeghidour, Neil},
  booktitle = 	 {Proceedings of the 43rd International Conference on Machine Learning},
  pages = 	 {62108--62127},
  year = 	 {2026},
  editor = 	 {Zhang, Tong and Dudik, Miroslav and Jaggi, Martin and Agarwal, Alekh and Li, Sharon and Schuurmans, Dale and Zhu, Jerry and Berkenkamp, Felix and Dong, Hanze and Bietti, Alberto},
  volume = 	 {306},
  series = 	 {Proceedings of Machine Learning Research},
  month = 	 {06--11 Jul},
  publisher =    {PMLR},
  url = 	 {https://proceedings.mlr.press/v306/labiausse26a.html},
  }

@misc{an2026openstbench,
 title={{OpenSTBench}: Beyond Semantic Evaluation for Speech Translation},
 author={An, Yanjie and Zhao, Yuxiang and Zhang, Yichi and Zheng, Qixi and Tu, Yujie and Deng, Keqi and Yu, Kai and Chen, Xie},
 year={2026}, eprint={2605.30792}, archivePrefix={arXiv},
 url={https://arxiv.org/abs/2605.30792},
  primaryClass = {eess.AS}
}

@misc{deng2026duras2st,
 title={{DuraS2ST}: Chain-of-Thought and Reinforcement Learning for Duration-Aligned Speech-to-Speech Translation},
 author={Deng, Yayue and Wang, Dingdong and Hu, Yuxuan and Li, Jinyu and Liu, Yanqing and Wang, Yuanyuan and Chen, Weidong and Meng, Helen M. and Liu, Shujie and Wu, Xixin},
 year={2026}, eprint={2609.33742}, archivePrefix={arXiv}, primaryClass={cs.SD},
 url={https://arxiv.org/abs/2609.33742}
}

@misc{choi2026dials2st,
  title = {{DIAL-S2ST}: Learning Latency Control for Simultaneous Speech-to-Speech Translation},
  author = {Choi, Jeongsoo and Ahn, Junseok and Chung, Joon Son},
  year = {2026},
  eprint = {2610.32214},
  archivePrefix = {arXiv},
  primaryClass = {cs.CL},
  url = {https://arxiv.org/abs/2610.32214}
}

@misc{choi2026xdubbing,
  title = {{X-Dubbing}: Speech-to-Speech Translation under Temporal and Visual Constraints},
  author = {Choi, Jeongsoo and Kim, Ji-Hoon and Ahn, Junseok and Chung, Joon Son},
  year = {2026},
  eprint = {2610.32042},
  archivePrefix = {arXiv},
  primaryClass = {cs.CL},
  url = {https://arxiv.org/abs/2610.32042}
}

@inproceedings{camp2023mos,
  title = {{MOS vs. AB: Evaluating Text-to-Speech Systems Reliably Using Clustered Standard Errors}},
  author = {Joshua Camp and Tom Kenter and Lev Finkelstein and Rob Clark},
  year = {2023},
  booktitle = {Interspeech 2023},
  pages = {1090--1094},
  doi = {10.21437/Interspeech.2023-2014},
  url = {https://www.isca-archive.org/interspeech_2023/camp23_interspeech.html}
}

@article{cheng2024clasi,
    title = "Towards Achieving Human Parity on End-to-end Simultaneous Speech Translation via {LLM} Agent",
    author = "Cheng, Shanbo and Huang, Zhichao and Ko, Tom and Li, Hang and Peng, Ningxin and Xu, Lu and Zhang, Qini",
    journal = "arXiv preprint arXiv:2407.21646",
    year = "2024",
    doi = "10.48550/arXiv.2407.21646",
    url = "https://arxiv.org/abs/2407.21646"
}

@article{kano2024atd,
    title = "Average Token Delay: A Duration-aware Latency Metric for Simultaneous Translation",
    author = "Kano, Yasumasa and Sudoh, Katsuhito and Nakamura, Satoshi",
    journal = "Journal of Natural Language Processing",
    year = "2024",
    volume = "31",
    number = "3",
    pages = "1049--1075",
    doi = "10.5715/jnlp.31.1049",
    url = "https://www.jstage.jst.go.jp/article/jnlp/31/3/31_1049/_article/-char/en"
}
